\documentclass[aps,prc,tightenlines,twoside,
fleqn,superscriptaddress]{revtex4-2}

\usepackage{amsmath,amsthm,amssymb}
\usepackage{graphicx}
\usepackage{soul}
\usepackage{color}
\usepackage{xcolor}
\usepackage{array}
\usepackage{booktabs}
\usepackage{multirow}
\usepackage{comment}

\usepackage[colorlinks,
bookmarks=true,
linkcolor=blue,
urlcolor=blue,
anchorcolor=black,
citecolor=blue
]{hyperref}

\colorlet{darkgreen}{green!50!black}
\colorlet{brightyellow}{yellow!75!red}
\colorlet{orange}{red!50!yellow}
\colorlet{darkblue}{blue!60!black}
\colorlet{darkred}{red!80!black}

\newcommand{\tfo}{T_{\mathrm{fo}}}

\newcommand{\taufo}{\tau_{\mathrm{fo}}}

\begin{document}
	
\title{Parameterizing Smooth Viscous Fluid Dynamics With a Viscous Blast Wave}


\author{Zhidong Yang} 
\email{zdyang07@163.com}
\affiliation{School of Physics and Astronomy, Shanghai Key Laboratory for Particle Physics and Cosmology, Shanghai Jiao Tong University, Shanghai 200240, China}
\affiliation{Cyclotron Institute and Department of Physics and Astronomy, Texas A\&M University, College Station TX 77843, USA}

\author{Rainer J.\ Fries}
\email{rjfries@comp.tamu.edu}

\affiliation{Cyclotron Institute and Department of Physics and Astronomy, Texas A\&M University, College Station TX 77843, USA}

\date{\today}

\begin{abstract}
Blast wave fits are widely used in high energy nuclear collisions to capture essential features of global properties of systems near kinetic equilibrium. They usually provide temperature fields and collective velocity fields on a given hypersurface. We systematically compare blast wave fits of fluid dynamic simulations for Au+Au collisions at $\sqrt{s_{NN}}=200$ GeV and Pb+Pb collisions at $\sqrt{s_{NN}}=2.76$ TeV with the original simulations.
In particular, we investigate how faithful the viscous blast wave introduced in \cite{Yang:2022yxa} can reproduce the given temperature and specific shear viscosity fixed at freeze-out of a viscous fluid dynamic calculation, if the final spectrum and elliptic flow of several particle species are fitted. We find that viscous blast wave fits describe fluid dynamic pseudodata rather well and reproduce the specific shear viscosities to good accuracy.
However, extracted temperatures tend to be underpredicted, especially for peripheral collisions.
We investigate possible reasons for these deviations. We establish maps from true to fitted values. These maps can be used to improve raw fit results from viscous blast wave fits. 
Although our work is limited to two specific, albeit important, parameters and two collision systems, the same procedure can be easily generalized 
to other parameters and collision systems.

\end{abstract}


\maketitle

\section{Introduction}
\label{sec:intro}

Blast waves are simple and effective tools that provide snapshots of a system that is close enough to local kinetic equilibrium
so that macroscopic concepts like temperature, local collective velocity, and shear and bulk stress can be used.
The full dynamics of such systems is usually very well captured by viscous fluid dynamic equations of motion. If the equation of state 
and a sufficient number of transport coefficients (e.g. shear and bulk viscosity) of the system are known, fluid dynamics can evolve 
the system starting from a given initial state. In contrast, blast waves usually provide a static picture, typically the temperature 
field $T(x)$ and collective flow field $u^\mu(x)$ on a hypersurface $\Sigma$ consisting of points $r^\mu$. E.g., if the hypersurface 
of kinetic freeze-out in an expanding system is chosen, these fields can often be directly related to
observable particles of the system and can be obtained by fitting to data. 

In high energy nuclear collisions, blast waves have routinely been used to analyze 
the properties of the fireball of hadrons at the time of kinetic freeze-out \cite{Siemens:1978pb,Schnedermann:1993ws,Retiere:2003kf,Florkowski:2004tn,ALICE:2013mez,Sun:2014rda,Melo:2015wpa,Mazeliauskas:2019ifr,Liu:2022ikt}. 
After kinetic freeze-out hadrons free stream to the detectors and thus directly carry information about the freeze-out hypersurface. 
Blast wave parameters like the freeze-out temperature and average radial flow velocity can be determined by fits to transverse 
momentum spectra of hadrons. Additional information, like elliptic deformations of the fireball in coordinate and momentum space at the time of 
freeze-out can be extracted from fits of elliptic flow coefficients $v_2$. More recently, viscous corrections to blast waves have been 
considered \cite{Teaney:2003kp,Jaiswal:2015saa,Yang:2016rnw,Yang:2022yxa}. They can be used to extract the specific shear viscosity $\eta/s$ for hadronic matter 
at the freeze-out temperature \cite{Yang:2022yxa} and parton matter at the pseudocritical temperature \cite{Yang:2022ixy}. Viscous blast waves can also be used to extend the range of validity of ideal blast wave fits 
to larger transverse hadron momenta $P_T$. 

The question arises to what extent blast waves, which use certain simplifying approximations, can faithfully capture important properties of the full dynamical system.
The global picture is one of two successive approximations
\begin{equation}
   \text{real collision system} \succ \text{fluid dynamic simulation} \succ \text{blast wave fit}  \nonumber
\end{equation}
where $\succ$ means "approximated by". 
Viscous fluid dynamic simulations are widely accepted to give accurate descriptions of the
low transverse momentum region of high energy nuclear collisions, and analyses of experimental data with fluid dynamic
simulations continue to be an active field of study \cite{Gale:2013da,Bernhard:2019bmu,JETSCAPE:2020avt}. There are several approximations that enter when describing the evolution of the system and its freeze-out
with fluid dynamics. When dealing with the late hadronic phase the main issue is the instantaneous 
approximation made for the freeze-out. A sudden freeze-out corresponds to
the mean free path of hadrons suddenly rising to infinity, for all hadron species at once. Simulations
with hadronic transport models solving the Boltzmann equation show a more realistic picture of a gradual freeze-out process.
A systematic comparison of fluid dynamics and Boltzmann transport in the hadronic phase could illuminate uncertainties arising from these approximations. However, the quantification of these types of uncertainties is outside the scope of this work. 

Here, we focus on the second step in the approximation chain, and quantify uncertainties that arise when fluid dynamic systems are approximated by 
blast waves at freeze-out. One might ask whether the deployment of blast waves is still needed, given the proliferation 
of viscous fluid dynamic codes, and the cheap numerical cost, at least for 2+1D codes. The motivation lies only partly in 
the simplicity of blast wave fits. A second, important argument is their complementarity. Fluid dynamic calculations come
with their own set of uncertainties, many of which are not shared by blast waves. As an important example, fluid dynamics
computes flow fields using, among other input, initial conditions and an equation of state. The final flow field at freeze-out will depend on these inputs. On the other hand, blastwaves are independent of these specific inputs and rather find the final
flow field by fits to data. Of course, blast waves suffer from other, complementary, uncertainties which will be discussed
here. We refer the reader to our work \cite{Yang:2022yxa} for an example that uses the complementarity of blast waves to extract
properties of hadronic matter at freeze-out.

The outline of our work is as follows. We use smooth relativistic viscous fluid dynamics to create systems close to local 
equilibrium as they typically occur in the late stages of high energy nuclear collisions. Specifically, we utilize the 
viscous fluid dynamics code MUSIC \cite{Schenke:2010nt,Gale:2013da,Ryu:2015vwa} to generate simulation pseudodata.
The setups of the calculations roughly reflect conditions in Au+Au collisions at the Relativistic Hadron Collider (RHIC) and Pb+Pb collisions at the Large Hadron Collider (LHC), as described below, although a precise description of data is not the point of this work.
For direct comparisons of MUSIC simulations to data we refer the reader to \cite{Schenke:2010nt,Ryu:2015vwa}.
We subsequently use the viscous blastwave introduced by us in \cite{Yang:2022yxa} to fit the transverse momentum spectra 
and elliptic flow computed in MUSIC for several species of identified hadrons. This blast wave is a generalization of the 
ideal blast wave by Retiere and Lisa (RL) \cite{Retiere:2003kf}.
The blast wave is defined by an ansatz for the temperature, flow field and space-time structure 
of the freeze-out hypersurface, as discussed in detail below. The ansatz contains several parameters with physical meaning, like the
size of the fireball at freeze-out, and the shape of the collective flow field. We focus here on the extracted freeze-out temperature 
$\tfo$ and the specific shear viscosity $\eta/s$, i.e. the ratio of shear viscosity $\eta$ and entropy density $s$, at freeze-out. 
The extracted values can be directly compared to their "true" counterparts used in MUSIC. We also compare the flow field extracted by the blast wave 
fit to its counterpart in MUSIC.

Differences that are seen between "true" fluid dynamic results and extracted values can be due to simplifications made in the blast wave ansatz,
or due to limiations imposed on the range and error bars on pseudodata. We will discuss both of these below.
For future applications of viscous blast wave fits it is is important to understand and quantify the uncertainties and systematic biases in the fit results.
As an example, we introduce a map from the "true" values of $\tfo$ and $\eta/s$ set in fluid dynamic simulations to the corresponding 
values extracted from viscous blast wave fits of spectra and elliptic flow. The inverse map can be used to improve blast wave fits by 
systematic unfolding. Blast waves equipped with such procedures to remove systematic bias will have significantly improved precision. 
In this work we take a first step in this direction. Ref. \cite{Yang:2022yxa} in which $\eta/s$ in the hadronic phase is extracted from 
experimental data serves as an example for the usefulness of such procedures.

The paper is organized as follows. In section \ref{sec:bw}, we review the viscous Retiere-Lisa blast wave and discuss the approximations made. In section \ref{sec:hydro}, we describe the setup of the MUSIC hydrodynamic calculations and the pseudodata that is fitted. In section \ref{sec:fits}, we provide the results of the viscous blast wave fits. In section \ref{sec:analysis}, we discuss the relation between fluid dynamic parameters and blast wave fit parameters and quantify the deviations of blast wave fits.
We conclude with a discussion of our results and possible improvements.

\section{Fluid Dynamics and Viscous Blastwave}
\label{sec:bw}

In this section we briefly review some basic concepts shared by both fluid dynamics freeze-out and blast waves. We will then discuss the particular
ansatz for the viscous blast wave in \cite{Yang:2022yxa}, based on the work by Retiere and Lisa \cite{Retiere:2003kf}. 
For a system close enough to local kinetic equilibrium one can assign a local temperature field $T(r)$ and a flow field $u^\mu(r)$ to describe the temperature 
and collective motion as a function of position 4-vector $r^\mu$. The particle distribution in the local rest frame of a fluid cell can then be written as
\begin{equation}
  \label{eq:fdist}
  f(r,p) = f_{0}(r,p) + \delta f(r,p)
\end{equation}
where $f_0$ is the equilibrium Bose or Fermi-distribution as a function of particle momentum $p^\mu =(E,\mathbf p)$ for given chemical potential $\mu$ and local temperature $T$, 
\begin{equation}
  f_0(r,p) = \frac{1}{e^{(p\cdot  u(r) -\mu)/T}\mp 1}  \, , 
\end{equation}
and $\delta f$ is a small correction that accounts for the out-of-equilibrium behavior. We neglect chemical potentials in this study and set $\mu=0$ here, but note that
realistic chemical potentials for stable hadrons were used in \cite{Yang:2022yxa}. 
The general form of the correction term is \cite{Damodaran:2020qxx,Yang:2022yxa}
\begin{equation}
  \label{eq:deltaf}
  \delta f (r,p) = \frac{\eta}{s} \frac{\Gamma(6)}{\Gamma(4+\lambda)} \left(\frac{E}{T}\right)^{\lambda-2} 
  \frac{p_\mu p_\nu}{T^3}\sigma^{\mu\nu} f_{0}(r,p)  \, .
\end{equation}
Here the shear stress tensor $\pi^{\mu\nu}$ has been expressed by its Navier-Stokes approximation, $\pi^{\mu\nu} = 2 \eta \sigma^{\mu\nu}$, where
$\sigma^{\mu\nu}$ is the traceless gradient tensor, defined as
\begin{equation}
  \sigma^{\mu\nu} = \frac{1}{2} \left( \nabla^\mu u^\nu + \nabla^\nu u^\mu \right) - \frac{1}{3} \Delta^{\mu\nu} \nabla_\lambda u^\lambda  \, .
  \label{eq:sigma}
\end{equation}
We have used the notation $\nabla^\mu = \Delta^{\mu\nu} \partial_\nu$, with $\Delta^{\mu\nu} = g^{\mu\nu} - u^\mu u^\nu$, for the derivative perpendicular to the flow field vector $u^\mu$. In the following we will use the standard choice $\lambda=2$ for the residual momentum dependence of the correction, which is widely used in relativistic viscous hydrodynamics \cite{Song:2007ux} including MUSIC. In order to guarantee the applicability of the equations in this section, the correction term $\delta f$ needs to be small. We have checked that $\delta f(r,p)$ is less than 20\% of $f_0(r,p)$ for the majority of transverse momentum bins within the fit ranges, with very few bins receiving corrections up to 35\%.
These corrections are smaller than the necessary upper bound $\delta f/f_0 \lesssim 1$ \cite{JETSCAPE:2020mzn}.

When solving the viscous fluid dynamics equations of motion, numerical stability requires second order gradient terms to be included, leading to equations of motion for the shear stress tensor $\pi^{\mu\nu}$ and bulk stress $\Pi$ \cite{Israel:1979wp,Baier:2006um,Song:2007ux,Karpenko:2013wva,Romatschke:2017ejr}. At freeze-out, it is then convenient to compute $\delta f$ directly from the shear stress tensor $\pi^{\mu\nu}$. 
On the other hand, for the blast wave it is more practical to utilize the Navier-Stokes approximation $\pi^{\mu\nu} = 2 \eta \sigma^{\mu\nu}$ and to compute viscous corrections using Eq.\ (\ref{eq:deltaf}). In that case $\delta f$ is calculated simply from the flow field, which can be independently constrained by fits to flow data and the specific shear viscosity $\eta/s$ of nuclear matter.
The differences between the two approaches of calculating $\delta f$ at freeze-out, $\pi^{\mu\nu}$ vs Navier-Stokes, are parametrically small in situations of small gradients towards the end of the time evolution. However, they could still be noticeable at freeze-out in realistic systems and are part of the uncertainties to be accounted for.

In both blast wave and fluid dynamic freeze-out, the invariant particle momentum spectrum emitted from a hypersurface $\Sigma$ in Minkowski space is given by the Cooper-Frye formula 
\cite{Cooper:1974mv}
\begin{equation}
  \label{eq:cf}
  \frac{dN}{dY d^2P_T} = g \int\frac{p\cdot d\Sigma}{(2\pi)^3} f(r, p)
\end{equation}
where $g$ is the degeneracy factor for a given particle and $d\Sigma^\mu$ is the forward normal vector on the freeze-out hypersurface.
The momentum vector in the laboratory frame is written as usual, $p^\mu = (M_T \cosh Y, P_T \cos\psi,$ $P_T \sin\psi, M_T\sinh Y)$, in terms of the transverse momentum $P_T$, the longitudinal momentum rapidity $Y$ and the azimuthal angle $\psi$ in the transverse plane. $M_T^2 = P_T^2+M^2$ defines the transverse mass $M_T$ for a hadron of mass $M$. 
The final particle spectrum at freeze-out is usually calculated on a hypersurface at constant temperature $T=\tfo$. In contrast, in fluid dynamics this isothermal hypersurface, as well as the flow field $u^{\mu}$ on it can be computed in the simulation itself. For the blast wave we have to choose ans\"atze for both.

Following Ref.\ \cite{Retiere:2003kf} we assume that freeze-out from an isothermal hypersurface at temperature $T$ can be approximated by freeze-out from a hypersurface at constant proper longitudinal time $\tau$. We enforce boost invariance, which is a good approximation for nuclear collisions around midrapidity at top RHIC and LHC energies and is also often found in fluid dynamic calculations. To keep the blast wave simple we have to restrict ourselves to describing smooth fluid dynamics which corresponds to the averaging over many events. We can then assume that the hypersurface in the $x-y$-plane is approximately an ellipse with semi-axes $R_x$ and $R_y$ in $x$- and $y$-directions, respectively. We define the coordinate axes such that the impact parameter $b$ of the collision is measured along the $x$-axis. In the following we use the reduced radius $\rho=\sqrt{x^2/R_x^2+y^2/R_y^2}$ together with the azimuthal angle $\theta$, with $\tan\theta = (R_x y)/(R_y x)$, and space time rapidity $\eta_s = 1/2 \log[(t+z)/(t-z)]$ to carry out the integral over the hypersurface. Restricting ourselves to hadrons measured at midrapidity $Y=0$ and changing to convenient coordinates we
obtain the final expression for the particles from the blastwave:
\begin{equation}
  \frac{dN}{dY d^2P_T} = g \tau R_x R_y M_T \int_0^1 d\rho \int_0^{2\pi} d\theta \int_{-\infty}^\infty d\eta_s \frac{\rho \cosh\eta_s}{(2\pi)^3}  
   f_0(\rho,\theta,\eta_s; u\cdot p) \left[ 1+\frac{\eta}{s} \frac{1}{T^3} p^\mu p^\nu \sigma^{\mu\nu} \right]    \, .
  \label{eq:spectrum} 
\end{equation}

Next, we have to make an ansatz for the collective flow field. The general parameterization is 
\begin{equation}
  \label{eq:upara}
  u^\mu = \left( \cosh \eta_s \cosh \eta_T, \sinh \eta_T \cos\phi_u, \sinh\eta_T \sin\phi_u, \sinh\eta_s \cosh\eta_T
  \right)
\end{equation}
where $\eta_T$ is the transverse rapidity in the $x-y$-plane, and $\phi_u$ is the azimuthal angle of the flow
vector in the transverse plane. Boost invariance fixes the longitudinal flow rapidity to be equal
to the space time rapidity $\eta_s$. We follow Retiere and Lisa and choose to model the transverse flow velocity $v_T = \tanh \eta_T$
as \cite{Retiere:2003kf}
\begin{equation}
  v_T =  \rho^n \left(\alpha_0 + \alpha_2 \cos(2 \phi_u) \right)
\end{equation}
which encodes a Hubble-like velocity ordering with an additional shape parameter $n$. $\alpha_0$ is the average velocity on the 
boundary $\rho=1$, and $\alpha_2$ parameterizes the elliptic deformation of the flow field built up from the initial elliptic spatial deformation of the system. Flow vectors tend to be tilted towards the smaller axis of the ellipse. In the RL approach they are chosen to be
perpendicular to the elliptic surface at $\rho=1$, i.e.\ $\tan \phi_u = R_x^2/R_y^2 \tan \phi$, where $\phi=\arctan y/x$ is the azimuthal angle of the position $r^\mu$. 

With a parameterization of the flow field at hand, the next step is the calculation of the gradient tensor $\sigma^{\mu\nu}$. This has been carried out in \cite{Yang:2022yxa} and we refer the reader to the details in that reference. We want to point out that temporal derivatives are calculated using ideal fluid dynamic equations of motion rather than the free-streaming approximation \cite{Teaney:2003kp,Jaiswal:2015saa}. This introduces the nuclear matter equation of state, specifically the speed of sound squared $c_s^2$ into our calculation of $\delta f$.

The blast wave ansatz has several parameters which allow us to adjust the flow field and the hypersurface, as well as the temperature, specific
shear viscosity, and speed of sound squared at freeze-out. The full set of parameters
is $\tilde{\mathcal{P}} = (\taufo, R_x, R_y,\tfo, n, \alpha_0, \alpha_2, \eta/s, c_s^2)$. As in Ref.\ \cite{Yang:2022yxa} we drop $c_s^2$ from this list and rather use guidance on the hadronic matter equation of state from existing literature which gives $c_s^2=0.16\sim0.17$ for T = 110-140 MeV \cite{Teaney:2002aj}. We also use a simple geometric argument for the time-dependence of the system size along the impact vector, $R_x= (R_0-b/2)+ \tau c_\tau (\alpha_0+\alpha_2)$, 
to determine $R_x$. Here $R_0$ is the radius of the colliding nucleus, $b$ is the impact parameter and $c_\tau$ relates the time-averaged surface velocity to the final velocities $\alpha_0$ and $\alpha_2$. The value of $c_\tau$ can be inferred from typical radial velocity-vs-time curves obtained in fluid dynamic simulations \cite{Song:2007ux} and we set $c_\tau=0.65$ here. Note that the viscous blast wave depends on the parameters $\taufo$, $R_x$ and $R_y$ separately, and not just on the total volume $\tau R_x R_y$ and the elliptic deformation $R_y/R_x$, as is the case for the ideal blast wave. With $R_x$ fixed in each instance we are left with the $R_y/R_x$ as a fit parameter.

Despite the large number of parameters it is clear that the blast wave has introduced significant simplifications compared to the freeze-out calculated 
in fluid dynamics. They come mostly from the simplified shape of the hypersurface (constant proper time) and the spatial structure of the flow field.
Two more major approximations are made for sake of simplicity. First, resonance decays are usually neglected in blast wave calculations, and only hadrons stable under strong decays are taken into account. Secondly, correction terms to the particle distribution $f$ due to bulk stress have been ignored. They could in principle be added and we plan to do so in the future. We summarize the five major approximations compared to fluid dynamic freeze-out in the following list:
\begin{itemize}
\item Navier-Stokes approximation used for $\delta f$.
\item Certain aspects of the shape of the hypersurface are fixed.
\item General shape of the flow field is fixed.
\item Lack of resonance production and decay.
\item Missing bulk stress effects on particle distributions.
\end{itemize}

Since we have eliminated event-by-event fluctuations from the comparison, the effects of event-by-event fluid dynamic simulations compared to smooth fluid dynamics
need to be considered separately. They are not included in the study below. The same is true for deviations of state-of-the-art 3+1D fluid dynamics from the boost-invariant 2+1D fluid dynamics
used here.
The effects of fluctuations and breaking of boost invariance have already been studied within fluid dynamics \cite{Schenke:2010rr,Qiu:2011iv} and can be added to the considerations in this work.

\section{Generation of MUSIC Pseudodata}
\label{sec:hydro}

We use the viscous hydrodynamics code MUSIC to simulate averaged nuclear collisions at RHIC and LHC energies at various impact parameters. MUSIC is a  relativistic second-order viscous hydrodynamics code for heavy ion collisions \cite{Schenke:2010nt,Schenke:2010rr,Ryu:2015vwa}. We choose boost-invariant (2+1)D mode consistent with the boost-invariant blast wave set-up.
We use the built-in optical Glauber model to generate initial conditions with the appropriate nucleon-nucleon cross section and an overall normalization roughly consistent with pertinent multiplicity data for Au+Au collisions at
RHIC at $\sqrt{s_{NN}}=200$ GeV and Pb+Pb collisions at the LHC at $\sqrt{s_{NN}}=2.76$ TeV. Other
collision systems can be treated similarly.
We use the equation of state (EOS) s95p-v1.2 in MUSIC, and the default MUSIC bulk viscosity. The shear viscosity over entropy ratio $\eta/s$ is chosen to be a constant which we vary as a parameter. We freeze out at pre-determined temperatures $\tfo$ and compute the final spectra and elliptic flow for pion, kaons and protons, including resonance decays and including viscous corrections to freeze-out. 
The detailed MUSIC settings are documented in Appendix \ref{sec:app1}.

Recall that we want to establish a map from the temperature $\tfo^{(\text{extr})}$ and specific shear viscosity $(\eta/s)^{(\text{extr})}$ extracted from a blast wave fit of the pseudodata to the true values $\tfo^{(\text{true})}$ and $(\eta/s)^{(\text{true})}$ used in the generation of these pseudodata. 
To focus on the relevant region tested in heavy ion collisions, we choose nine points in the $\tfo^{(\text{true})}$-$(\eta/s)^{(\text{true})}$-plane for simulations at RHIC energy, such that the corresponding fitted values $(\tfo^{(\text{extr})},(\eta/s)^{(\text{extr})})$ are roughly consistent with the values extracted from RHIC data in Ref.\ \cite{Yang:2022yxa}, for some impact parameter $b$. Thus, a single set of parameters to run MUSIC  consists of an impact parameter $b$ and values $\tfo^{(\text{true})}$, $(\eta/s)^{(\text{true})}$. We run MUSIC and perform a blast wave fit to the resulting hadron spectra and elliptic flow for all nine such sets at RHIC energies. Similarly, we choose eight points for Pb+Pb collisions at LHC.
The 9+8 sets of parameters are shown in Tab.\ \ref{tab:event}. We will refer to these points often as Set I.

\begin{table}[tbh]
	\centering
		\caption{\label{tab:event} ($\tfo$, $\eta/s$) from Set I chosen for MUSIC simulations of Au+Au and Pb+Pb collisions, 
 		together with the corresponding impact parameters $b$.} 
	\begin{tabular}{|c||c|c|c|c|c|c|c|c|c|}
		\hline
		$b$ (fm) (Au+Au) & 5 & 5 & 6 & 6.5 & 7 & 8 & 9 & 10.5 & 10.5\\
		\hline
		$b$ (fm) (Pb+Pb) &   & 5.3 & 6.3 & 6.9 & 7.4 & 8.5 & 9.6 & 11.1 & 11.1\\
		\hline
		$\tfo^{(\text{true})}$ (MeV)& 105 & 110 & 115 & 120 & 125 & 130 &  135 & 140 & 145 \\
		\hline
		$4\pi(\eta/s)^{(\text{true})}$ &6.03&  5.28& 4.52& 3.77& 3.02& 2.51&  2.01& 1.51& 1.01\\
		\hline
	\end{tabular}
\end{table}

\begin{table}[tbh]
	\centering
	\caption{\label{tab:range1} Preferred fit ranges for the Au+Au pseudodata. Similar fit ranges are used for Pb+Pb pseudodata.}
	\begin{tabular}{|c||l|l|l|l|l|l|}
		\hline
		\multirow{2}*{$\tfo^{(\text{true})}$ (MeV)} &\multicolumn{3}{c}{$P_T$-range spectra (GeV/c)} & \multicolumn{3}{|c|}{$P_T$-range $v_2$ (GeV/c)} \\
		\cline{2-7}
		& pion & kaon & proton & pion & kaon & proton  \\
		\hline
		\hline
		105 & 0.34-1.95 & 0.34-2.23 & 0.76-2.52 & 0.53-3.0 & 0.34-3.0 & 0.34-3.0 \\
		\hline
		110 & 0.34-2.37 & 0.34-2.68 & 0.34-3.0 & 0.34-3.0 & 0.34-3.0 & 0.34-3.0 \\
		\hline
		115 & 0.34-1.95 & 0.34-2.23 & 0.34-2.52 & 0.34-3.0 & 0.34-3.0 & 0.34-3.0 \\
		\hline
		120 & 0.40-1.95 & 0.40-2.09 & 0.34-2.37 & 0.34-2.84 & 0.34-3.0 & 0.34-3.0 \\
		\hline
		125 & 0.40-1.95 & 0.40-2.09 & 0.34-2.37 & 0.34-2.68 & 0.34-2.84 & 0.34-3.0 \\
		\hline
		\hline
		130 & 0.46-1.95 & 0.40-2.09 & 0.29-2.23 & 0.34-2.68 & 0.34-2.68 & 0.34-2.84 \\
		\hline
		135 & 0.34-1.82 & 0.29-1.95 & 0.24-2.09 & 0.34-2.52 & 0.34-2.68 & 0.34-2.68 \\
		\hline
		140 & 0.24-1.57 & 0.20-1.69 & 0.20-1.82 & 0.24-2.23 & 0.53-2.23 & 0.20-2.37 \\
		\hline
		145 & 0.24-1.57 & 0.20-1.69 & 0.20-1.82 & 0.24-2.23 & 0.53-2.23 & 0.20-2.37 \\
		\hline
	\end{tabular}
\end{table}  

The $P_T$-range of the pseudodata generated by MUSIC consists of the interval from 0 to 3 GeV/$c$. However, we have to restrict the range of data
in which we fit the blast wave to pseudodata. At very low momenta hadron spectra tend to be dominated by resonance decays, which are not included
in the blast wave. At momenta which are too high, hadron production receives viscous corrections larger than what can be reliably described by the Navier-Stokes approximation. Such restrictions of fit ranges for blast waves seem to be good practice in the literature. They are also used in Ref.\ \cite{Yang:2022yxa}. The fit ranges used for blast wave fits of the pseudodata here are shown in Tab.\ \ref{tab:range1}. They are inspired by what 
was used for good quality fits of experimental data in \cite{Yang:2022yxa}.

The fluid dynamic simulations do not provide useful uncertainty estimates on the pseudodata. For this study we choose uncertainties in line with error bars in the pertinent available experimental data. We assign 5\% uncertainty and 2\% uncertainty to pseudodata spectra and $v_2$, respectively. We add a pedestal of 0.002 to the uncertainty of $v_2$ for realistic error bars at smaller $P_T$ where $v_2$ is very small. 
The choices of fit ranges and error bars introduce additional uncertainties in our analysis. The dependence of blast wave fits on the choice of 
fit ranges were studied in Ref.\ \cite{Yang:2022yxa}. The additional uncertainty due to the choice of error bars for the pseudodata is studied
later in this work by varying the size of the assumed error bars, see Tab.\ \ref{tab:uncert}. 

We utilize Bayesian inference to extract likelihoods for the relevant parameters. We use the statistical analysis package from the Models and Data Analysis Initiative (MADAI) project \cite{MADAI:2013,Bernhard:2016tnd}. The MADAI package includes a Gaussian process emulator and a Markov Chain Monte Carlo. We use $N=500$ training points for each Gaussian emulator. Closure tests find the errors in the Gaussian emulator to be negligible compared to the assumed uncertainties in the pseudodata.

\begin{figure}[tbh]
	\centering
	\includegraphics[height=5.4 in,width=0.95\textwidth]{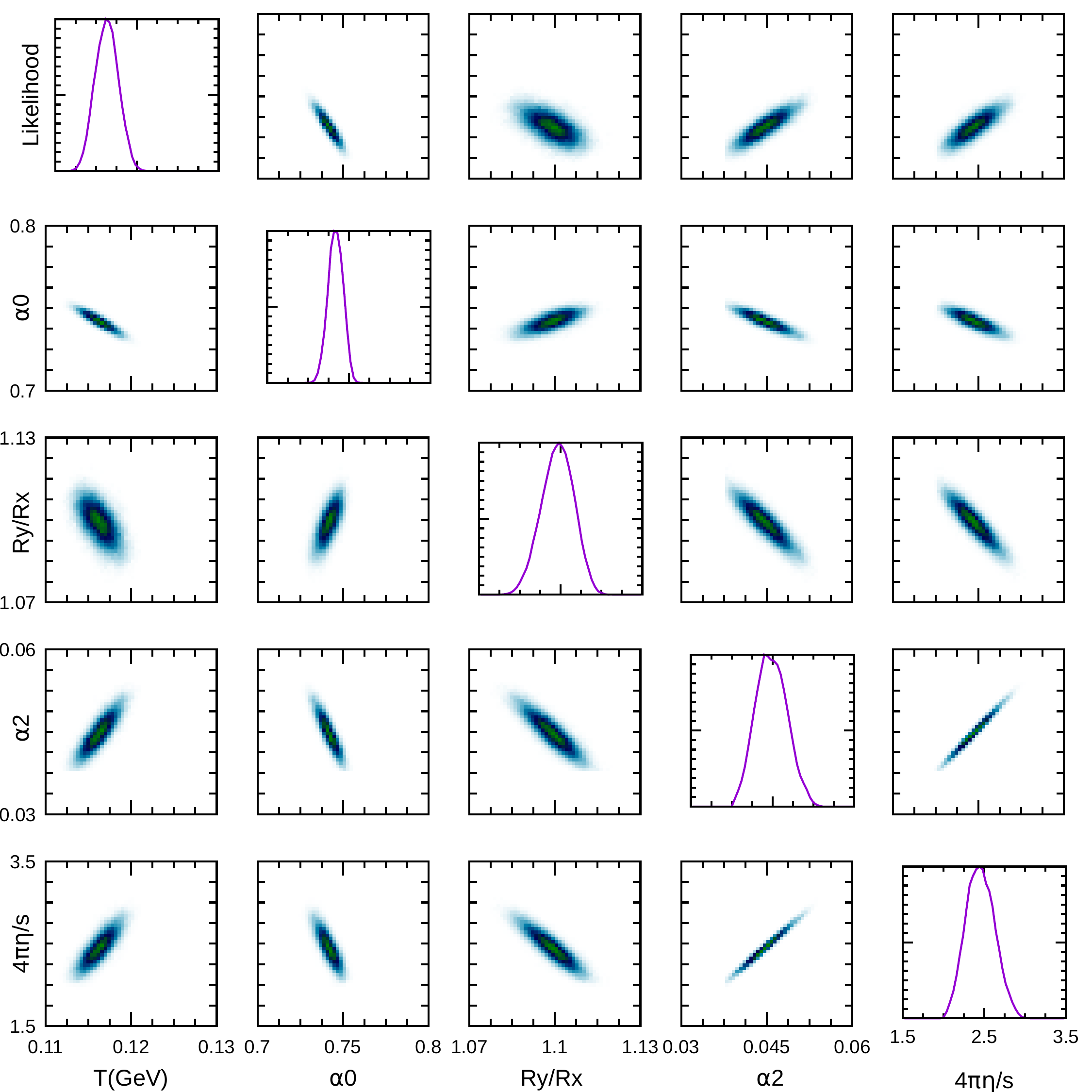}
	\caption{\label{fig:madai} Likelihood analysis for the MUSIC run for Au+Au with $\tfo^{(\text{true})}=130$ MeV, $(\eta/s)^{(\text{true})}=2.51/(4\pi)$.
		The axes show (from left to right, and top to bottom): $\tfo$ (GeV), $\alpha_0$, $R_y/R_x$, $\alpha_2$, $\eta/s$. The diagonal plots show the posterior likelihood distributions. The off-diagonal plots show correlations between parameters.} 
\end{figure}

\section{Blast Wave Fits}
\label{sec:fits}

We fit the following, reduced set of blast wave parameters in the Bayesian analysis: $\mathcal{P}=(\tfo, R_y/R_x, \alpha_0, \alpha_2, \eta/s)$. 
Chemical potentials $\mu$ are set to zero in MUSIC and in the blast wave. We will rather fit a normalized $P_T$-spectrum and do not utilize the absolute yield of hadrons
to reduce complexity. To further simplify the analysis we fix the radial shape parameters $n$ and the freeze-out time $\tau$ by choosing values close to those extracted 
from RHIC and LHC data in Ref.\ \cite{Yang:2022yxa}, see Tab.\ \ref{tab:resultau} and Tab.\ \ref{tab:resultpb}. 

\begin{table}[tbh]
	\centering
	\caption{\label{tab:resultau} Extracted parameter values $\mathcal{P}$ for MUSIC Au+Au collisions, together
	with values set for $\tau$ and $n$.}
	\begin{tabular}{|c|c||c|c|c|c|c|c|c|}
     \hline \multicolumn{2}{|c||}{Hydro Au+Au} & \multicolumn{7}{c|}{Blast Wave}  \\
		\hline
		$\tfo^{(\text{true})}$ & $4\pi (\eta/s)^{(\text{true})}$ & $\tfo $ (MeV) & $\alpha_0/c$ & $R_y/R_x$ & $\alpha_2/c$ & $4\pi \eta/s$ & $\tau$ (fm/$c$) & $n$ \\
		\hline
		\hline
		105 & 6.03 & 111.2 & 0.824 & 0.99 & 0.021 & 5.83 & 12.2 & 0.86 \\
		\hline
		110 & 5.28 & 114.0 & 0.822 & 1.01 & 0.021 & 5.43 & 11.4 & 0.87 \\
		\hline
		115 & 4.52 & 112.7 & 0.833 & 1.04 & 0.025 & 4.67 & 10.6 & 0.81 \\
		\hline
		120 & 3.77 & 113.9 & 0.820 & 1.06 & 0.028 & 3.75 & 9.8 & 0.84 \\
		\hline
		125 & 3.02 & 117.7 & 0.786  & 1.08 & 0.037 & 3.01 & 9.1 & 0.88 \\
		\hline
		\hline
		130 & 2.51 & 116.4 & 0.742  & 1.10 & 0.045 & 2.47 & 8.4 & 0.88\\
		\hline
		135 & 2.01 & 120.0 & 0.715  & 1.15 & 0.059 & 2.07 & 7.8 & 0.92\\
		\hline
		140 & 1.51 & 123.0 & 0.654 &  1.27 & 0.069 & 1.55  & 7.2 & 0.96\\
		\hline
		145 & 1.01 & 126.3 & 0.604 & 1.35 & 0.080 & 1.23 & 6.8 &  1.00 \\
		\hline
	\end{tabular}
\end{table}

\begin{table}[tbh]
	\centering
	\caption{\label{tab:resultpb} The same as Tab.\ \ref{tab:resultau} for Pb+Pb collisions. }
	\begin{tabular}{|l|l||l|l|l|l|l|l|l|}
		\hline
		 \multicolumn{2}{|c||}{Hydro Pb+Pb} & \multicolumn{7}{c|}{Blast Wave}  \\
		 \hline
		 $\tfo^{(\text{true})}$ & $4\pi (\eta/s)^{(\text{true})}$ & $\tfo$ (MeV) & $\alpha_0/c$ & $R_y/R_x$ & $\alpha_2/c$ & $4\pi \eta/s$ & $\tau$ (fm/$c$) & $n$ \\
		\hline
		\hline
		110 & 5.28 & 111.4 & 0.822  & 0.99 & 0.020 & 5.74 & 13.2 & 0.84\\
		\hline
		115  & 4.52 &117.5 & 0.827 & 1.00 & 0.023 & 4.73 & 12.6  & 0.87\\
		\hline
		120 & 3.77 & 120.0 & 0.818  & 1.03 & 0.026 & 3.98 & 12 & 0.85\\
		\hline
		125  &  3.02 &122.3 & 0.822  & 1.07 & 0.032 & 3.34 & 11.6 & 0.88\\
		\hline
		\hline
		130  & 2.51 &123.7 & 0.787  & 1.09 & 0.043 & 2.62 & 10.8 & 0.90\\
		\hline
		135  & 2.01 &125.6 & 0.750 & 1.13 & 0.054 & 2.01 & 10.0 & 0.94\\
		\hline
		140 & 1.51 &127.2 & 0.689  & 1.19 & 0.063 & 1.48  & 9.2 & 0.98\\
		\hline
		145  & 1.01 &130.3 & 0.642 & 1.24 & 0.075 & 1.18 & 8.6  & 1.00\\
		\hline
	\end{tabular}
\end{table}

As an example, we take a look at the case of the input parameter set $\tfo^{(\text{true})}=130$ MeV, 
$(\eta/s)^{(\text{true})}=2.51/(4\pi)$ in Au+Au collisions. Fig.\ \ref{fig:madai} shows posterior distributions and correlations
of the values $\mathcal{P}$ extracted from the Bayesian analysis. 
The likelihoods for all parameters 
exhibit well defined peaks. 
The preferred values (defined as the means) for the set $\mathcal{P}$ of parameters in this case are $\tfo=116.4$ MeV, $\alpha_0$=0.74$c$, $R_y/R_x$=1.10, $\alpha_2$=0.045$c$, $\eta/s$=2.5/4$\pi$, with fixed parameters $\tau=8.4$ fm/$c$ and $n$=0.88. 

\begin{figure}[tbh]
	\begin{tabular}{c}
		\begin{minipage}{0.495\textwidth}
			\includegraphics[width=1\textwidth]{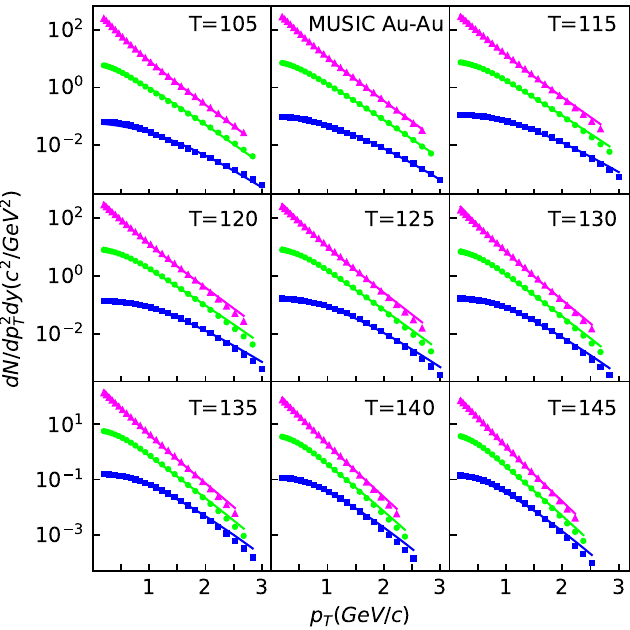}
		\end{minipage}
		
		\begin{minipage}{0.495 \textwidth}
			\includegraphics[width=1\textwidth]{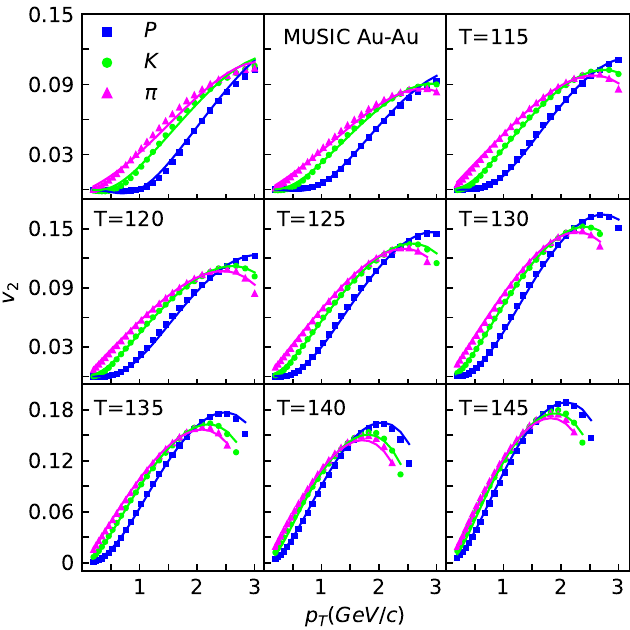}
		\end{minipage}
	\end{tabular}
	\caption{\label{fig:musicau} Transverse momentum spectra (right panel) and elliptic flows (left panel)  for protons, pions and kaons in Au+Au collisions calculated in MUSIC (symbols) together with blast wave calculations (solid lines) using the extracted parameter values in Tab.\ \ref{tab:resultau}. The nine plots correspond to the nine points of freeze-out temperature and specific shear viscosity given in Tab.\ \ref{tab:event}.}
\end{figure}

\begin{figure}[tbh]
	\begin{tabular}{c}
		\begin{minipage}{0.495\textwidth}
			\includegraphics[width=1\textwidth]{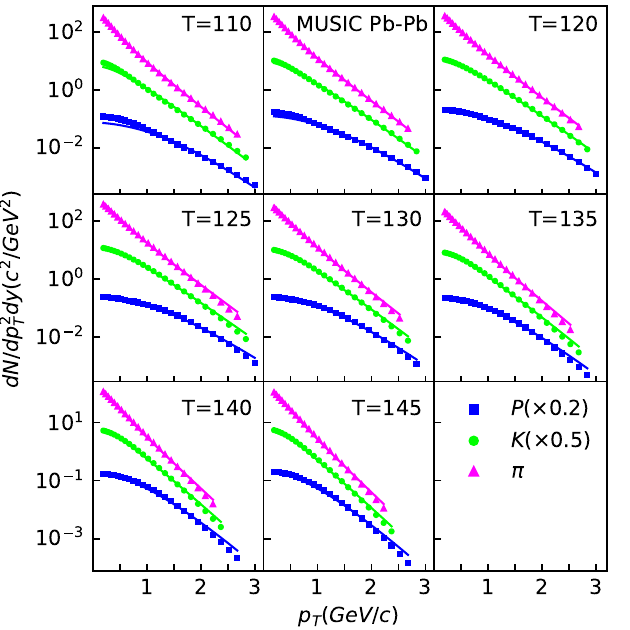}
		\end{minipage}
		\begin{minipage}{0.495 \textwidth}
			\includegraphics[width=1\textwidth]{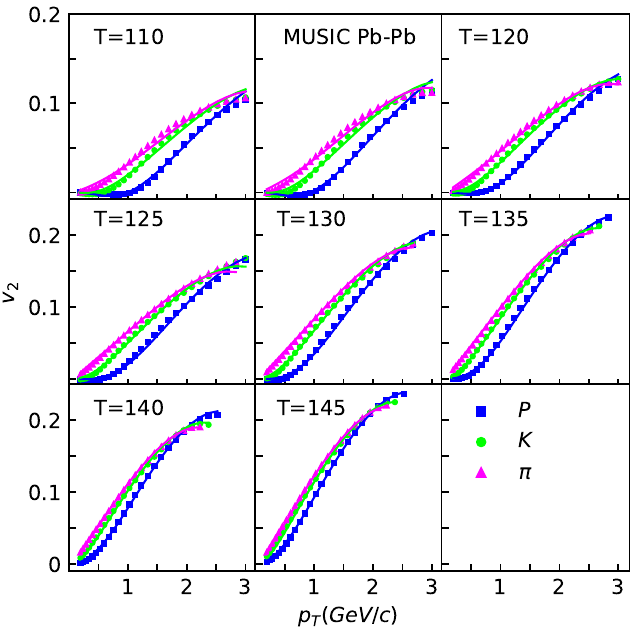}
		\end{minipage}
	\end{tabular}
	\caption{\label{fig:musicpb} Same as Fig.\ \ref{fig:musicau} for MUSIC Pb+Pb simulations.}
\end{figure}

We proceed analogously for the other $\tfo^{(\text{true})}$, $(\eta/s)^{(\text{true})}$ Au+Au points, and for the Pb+Pb points from Set I. The results are summarized in Tabs.\ \ref{tab:resultau} and \ref{tab:resultpb}. Let us first discuss the quality of the fit results for the observables.
Figs.\ \ref{fig:musicau} and \ref{fig:musicpb} show the identified spectra and $v_2$ computed from the blastwave with the
preferred values together with the MUSIC pseudodata for all impact parameters in Au+Au and Pb+Pb. The figures demonstrate that the fits to identified 
hadron spectra and elliptic flow are working quite well across the parameters chosen for Set I. 

In this work, the declared measures for success are the accuracies of the extracted specific shear viscosity and temperature compared to the true values
used in the simulations. For our example set $\tfo^{(\text{true})}=130$ MeV, $(\eta/s)^{(\text{true})}=2.51/(4\pi)$ in Au+Au the extracted specific shear viscosity
reflects the true value quite accurately --- 2.47 vs 2.51 --- while the extracted freeze-out temperature of 116 MeV is lower than the true value of 130 MeV. Looking across the entire set of fitted values we note that agreement between the extracted $\eta/s$ and the true value stays within 10\%, and often much closer, in all cases except
for the highest freeze-out temperatures. In contrast, there is a noteworthy trend for the extracted freeze-out temperatures. The agreement between true and 
extracted values is rather good for central collisions, in particular for Pb+Pb, and accuracy decreases toward peripheral collisions. For the latter, the extracted temperatures are underestimated by up to 20 MeV.

\begin{figure}[tb]
	\begin{tabular}{c}
		\begin{minipage}{0.395\textwidth}
			\includegraphics[width=1\textwidth]{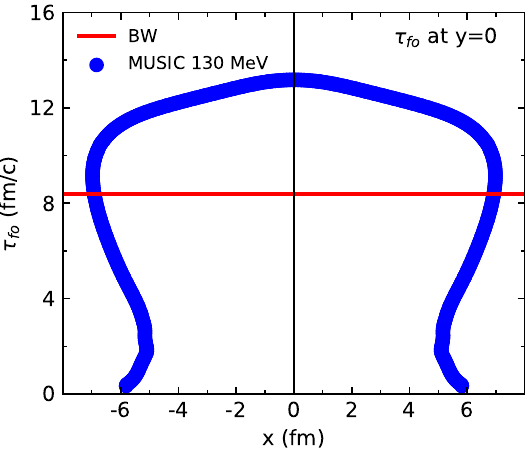}
		\end{minipage}
		
		\begin{minipage}{0.395 \textwidth}
			\includegraphics[width=1\textwidth]{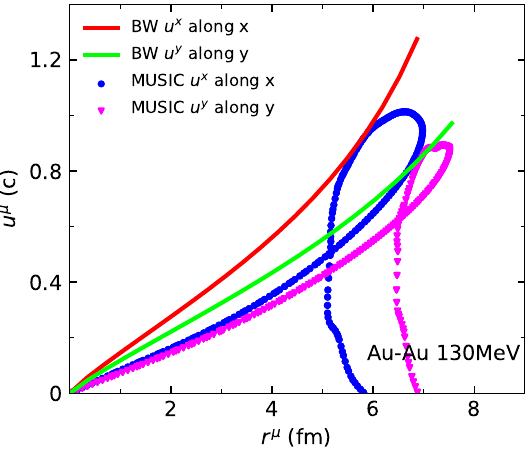}
		\end{minipage}
	\end{tabular}
	\caption{\label{fig:comparison} Left panel: Freeze-out time $\taufo$ and $x$-position for cells in MUSIC
for Au+Au collisions with $b=8$ fm and $\tfo=130$ MeV, together with the constant freeze-out time in the corresponding
 blast wave fit (red line). Right panel: The flow velocity component $u^x$ as a function of coordinate $x$ for freeze-out cells
at $y=0$ and $u^y$ as a function of coordinate $y$ for freeze-out cells
at $x=0$, for fluid cells in MUSIC in the same collision system. $u^x(x)$ and $u^y(y)$ in the corresponding blast wave fit 
are shown by the solid lines.
}
\end{figure}

To understand this behavior further it is informative to compare some details of the two simulated fireballs. The left panel of Fig.\ \ref{fig:comparison} shows the freeze-out 
cells in fluid dynamics vs the corresponding freeze-out hypersurface in the blast wave in the $t-x$-plane in Minkowski space. The "muffin" shape of the $T=\tfo$
hypersurface in fluid dynamics, which emerges from the competition of cooling and radial flow, is well known. On the other hand,
the choice made in blastwaves is to use an average constant time hypersurface. This comparison makes it clear that fields from
fluid dynamics and blast waves should not be expected to line up point-by-point. Rather, what one can reasonably hope for is for a matching of global
features and of the relevant numerical averages.

With that in mind we move to a comparison of components of the flow velocites $u^\mu(r)$
at freeze-out. $u^x(x)$ for $y=0$ and $u^y(y)$ for $x=0$ for our example set $\tfo^{(\text{true})}=130$ MeV, $(\eta/s)^{(\text{true})}=2.51/(4\pi)$ in Au+Au  
are shown for fluid dynamic cells at freeze-out and for the viscous blast wave in the right panel of Fig.\ \ref{fig:comparison}. We make note of two observations: 
The sizes $R_x$ and $R_y$ of the fireballs in $x$- and $y$-directions, respectively, indicated by the endpoints of the blast wave curves and the rightmost fluid cells, 
match up well.  On the other hand, it looks like the blastwave overestimates the radial flow in both directions compared to fluid dynamics. However, the weights 
with which the different fluid  cells contribute to the observed particle yields need to be taken into account. The sequences of MUSIC fluid cells fold in on 
themselves at large values of $x$ 
and $y$ because of the muffin shape of the hypersurface. Eq.\ (\ref{eq:spectrum}) exhibits a Jacobian $\tau\rho$ which makes the outer 
part of the muffin edge the largest contributer to particle production. This is where the cells with the largest radial flow 
are located, closer to their counterpart in the blast wave. After accounting for this effect $\langle u^x\rangle \approx 0.65$
for the freeze-out cells along the $x$-axis in fluid dynamics, while the corresponding value for the blast wave is 
$\langle u^x\rangle \approx 0.76$. The full expression in Eq.\ (\ref{eq:spectrum}) is used for the averaging here, with
the restriction $\eta=0$, $\theta=0$. Thus a residual mismatch remains, with the blast wave overestimating the radial flow. 
This is consistent with the lower fitted temperatures in the blast wave. For momenta which are not too small
the main effect of flow is an effective blue-shift of the temperature which is also reflected by the anti-correlation between
$T$ and $\alpha_0$ in Fig.\ \ref{fig:madai}. This explains why the pseudodata for this set can be fit rather well in the chosen range, although the extracted temperature 
is too low.

\begin{figure}[tbh]
			\includegraphics[width=0.4\textwidth]{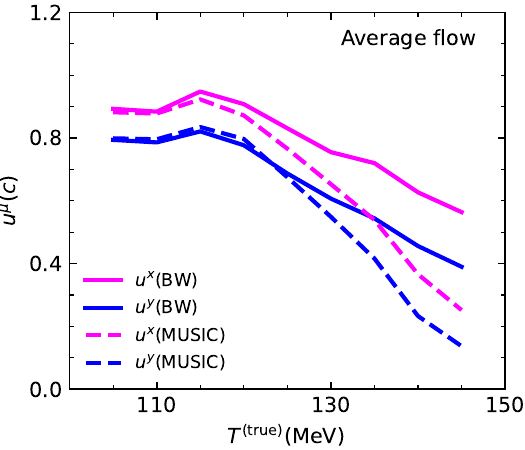}
		
	\caption{\label{fig:comparison2} The average radial flow velocities $\langle u^x\rangle$ and $\langle u^y\rangle$ computed along lines $y=0$ and $x=0$ as discussed
  in the text, using both MUSIC fluid cells and the blastwave with the corresponding extracted parameter values. The analysis is done for all sets of Au+Au input parameters
  and shown as a function of  $\tfo^{(\text{true})}$ of each set.
}
\end{figure}

The effect of lower temperatures being compensated by larger radial flow is largest in peripheral collisions and dissappears in more central collisions. 
This is confirmed by Fig.\ \ref{fig:comparison2} which shows the average flow values $\langle u^x\rangle$ and $\langle u^y\rangle$ along the $x$- and $y$-axes, 
respectively, discussed earlier for the example set, for all sets in Au+Au as a function of the temperature used in the MUSIC simulation. 
The highest temperatures, corresponding to the most peripheral collisions, see the largest deviation between fluid dynamics and blastwave fits for both radial
flow and the extracted temperature.
It is not quite clear why the peripheral blastwave fits prefer to trade a lower temperature against a higher flow field.  One can speculate that the simplistic 
$\tau$-dependence of the hypersurface in the blastwave obviously becomes a worse approximation in peripheral collisions.
On a practical level, a better separation of flow and temperature effects on observables, and thus better fits for the temperature, could presumably be achieved 
by employing a wider fit range, in particular through the inclusion of smaller momenta. To this end, resonance decays would have to be 
taken into account for the viscous blast wave, which is possible but beyond the scope of the current model. Here, we will simply document the 
biases in the fits and prepare the tools to remove them.

\begin{figure}[tbh]
	\begin{tabular}{c}
		\begin{minipage}{0.395\textwidth}
			\includegraphics[width=1\textwidth]{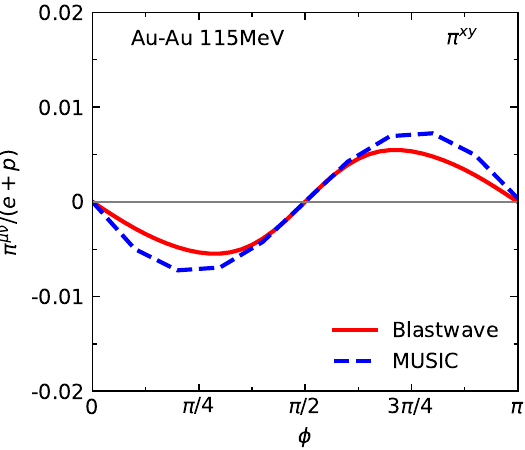}
		\end{minipage}
		
		\begin{minipage}{0.395 \textwidth}
			\includegraphics[width=1\textwidth]{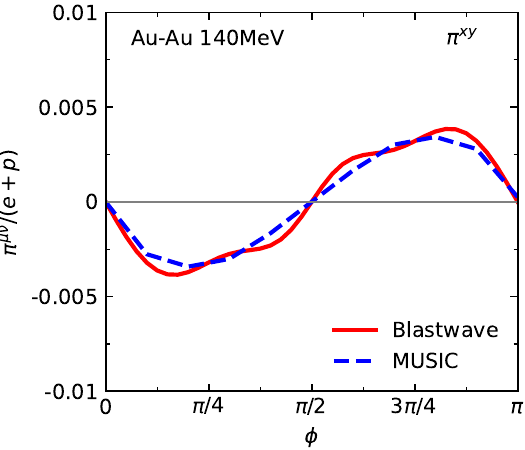}
		\end{minipage}
	\end{tabular}
	\caption{\label{fig:comparison3} Average shear stress tensor component $\pi^{xy}$ divided by enthalpy $e+p$ computed in the blastwave model with preferred values 
   and extracted from the MUSIC fluid dynamics hypersurface for Au+Au collisions from the sets with $\tfo=115$ MeV (left panel) and 140 MeV (right panel),
   as functions of the azimuthal angle $\phi$ in space-time. The average over the fireball is weighted with the contribution to the spectrum given by Eq.\ (\ref{eq:spectrum}).}
\end{figure}

Before we attempt to correct the extracted temperatures let us briefly look at the shear stress in both MUSIC and the blastwave. In Fig.\ \ref{fig:comparison3}
we compute the off-diagonal component $\pi^{xy}$ divided by the enthalpy $e+p$ as a function of the space-time angle $\phi$, averaged over the remaining coordinates of
the hypersurface for both MUSIC fluid dynamics and the blastwave for the preferred parameters. We show the results for Au+Au events at 
$\tfo^{(\text{true})}=115$ MeV and $140$ MeV, corresponding to rather central and peripheral collisions, respectively. Interestingly, the average shear stress in the fluid 
dynamic simulation is represented quite well by the viscous blast wave for both centralities. This comparison provides context for the good agreement we find for the
extracted values of the specific shear viscosity.

We summarize the results of this section in Figs.\ \ref{fig:errau} and \ref{fig:errpb} for Au+Au and Pb+Pb collisions, respectively.
The plots show the correlation between true and extracted values of our two main observables, the freeze-out temperature (right panels) and specific shear viscosity
(left panels). Diagonal lines for perfect correlation ("y=x") are added to guide the eye. The uncertainties for each point are combined uncertainties from the 
posterior likelihoods from the Bayesian analysis, and systematic uncertainties, see 
Appendix \ref{sec:app2}. 

\section{Correcting Blast Wave Bias}
\label{sec:analysis}

\begin{figure}[tb]
	\centering
	\begin{minipage}{0.4 \textwidth}
		\includegraphics[width=1\textwidth]{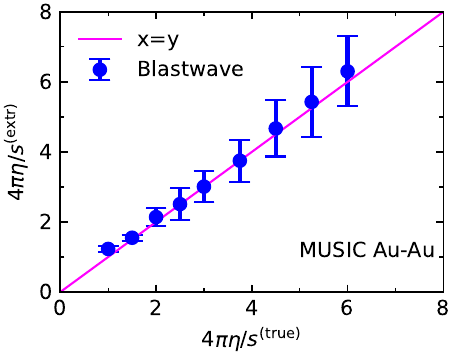}
	\end{minipage}
	\begin{minipage}{0.4 \textwidth}
		\includegraphics[width=1\textwidth]{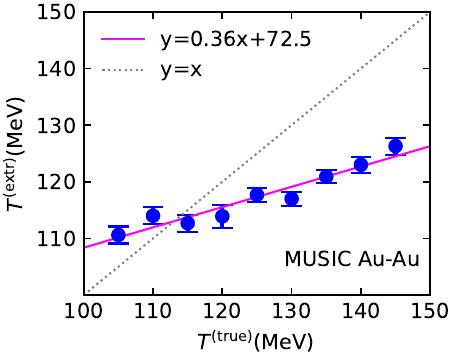}
	\end{minipage}
	\caption{\label{fig:errau}
		 Correlations between extracted and true freeze-out temperatures (right panel) and specific shear viscosities (left panel) from blast wave fits for Au+Au parameters
     from Set I.  The solid line in the right panel denotes the linear map discussed in the text.}
\end{figure}

\begin{figure}[tb]
	\centering
	\begin{minipage}{0.4 \textwidth}
		\includegraphics[width=1\textwidth]{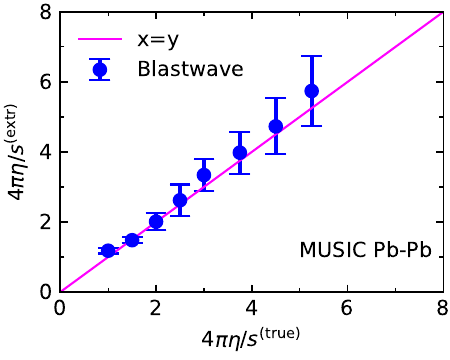}
	\end{minipage}
	\begin{minipage}{0.4 \textwidth}
		\includegraphics[width=1\textwidth]{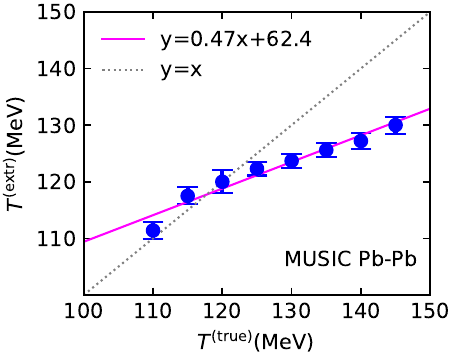}
	\end{minipage}
	\caption{\label{fig:errpb}
		The same as Fig.\ \ref{fig:errau} for the Pb+Pb case.}
\end{figure}

We now interpret the extracted values of freeze-out temperatures and specific shear viscosities as images of the original values 
$\tfo^{(\text{true})}$ and $(\eta/s)^{(\text{true})}$ under a map that is determined by the approximations made in the blast wave approach. 
The left panels of Figs.\ \ref{fig:errau} and \ref{fig:errpb} 
make it clear that --- within reasonable accuracy --- it is only necessary to consider the mapping of the true temperature onto the extracted temperature,
$\tfo^{(\text{true})} \mapsto \tfo^{(\text{extr})}$ and that it is a sufficiently accurate assumption that $(\eta/s)^{(\text{true})}\approx (\eta/s)^{(\text{extr})}$.
Moreover, Figs.\ \ref{fig:errau} and \ref{fig:errpb} tell us that a linear map should have the desired level of accuracy.

By minimizing
\begin{equation}
  \chi^2=\frac{1}{n}\sum_{i=1}^{n} \frac{ {\left( T_\mathrm{fo,i}^{  (\text{extr,lin})   }-T_\mathrm{fo,i}^{(\text{extr})} \right)}^2}{\sigma_i^2}
\end{equation}
for each input paramater point $i$, where $\sigma_i$ is the uncertainty of the extracted temperature at each point, we obtain an
approximation $T_\mathrm{fo}^{  (\text{extr,lin}) }$ to the extracted temperature by linear regression.
For Au+Au we obtain $T_\mathrm{fo}^{  (\text{extr,lin})} =0.36 T^{(\text{true})}_\mathrm{fo}  +72.5\, \text{MeV}$. 
Similarly, for Pb+Pb we have $\tfo^{(\text{extr,lin})}=0.47\, \tfo^{(\text{true})}+62.4\, \text{MeV}$. These linear functions are shown in
the right panels of Figs.\ \ref{fig:errau} and \ref{fig:errpb} in addition to the true and extracted values.
The linear maps provide a description of the extracted temperatures within the uncertainty bars. 
For all practical purposes we can thus identify $T_\mathrm{fo}^{  (\text{extr,lin}) }$ and $T_\mathrm{fo}^{  (\text{extr}) }$
for the rest of this section.

\begin{figure}[tb]
	\centering
	\begin{minipage}{0.4 \textwidth}
		\includegraphics[width=1\textwidth]{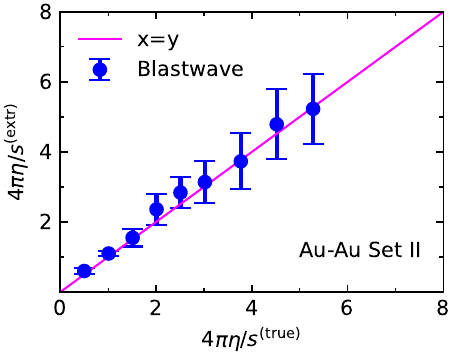}
	\end{minipage}
	\begin{minipage}{0.4 \textwidth}
		\includegraphics[width=1\textwidth]{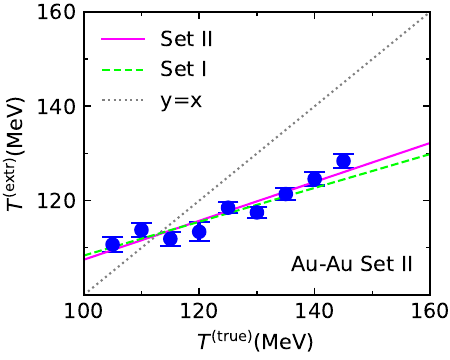}
	\end{minipage}
	\caption{\label{fig:errau2}
		The extracted freeze-out temperature (right panel) and specific shear viscosity (left panel) from blast wave fits (symbols) for MUSIC Au+Au simulations Set II. The solid line     
    is the linear map for this set obtained from linear regression, and the green dashed line is the map of Set I.}
\end{figure}

\begin{figure}[tb]
	\centering
	\begin{minipage}{0.4 \textwidth}
		\includegraphics[width=1\textwidth]{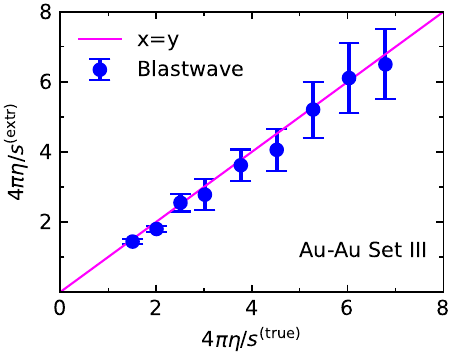}
	\end{minipage}
	\begin{minipage}{0.4 \textwidth}
		\includegraphics[width=1\textwidth]{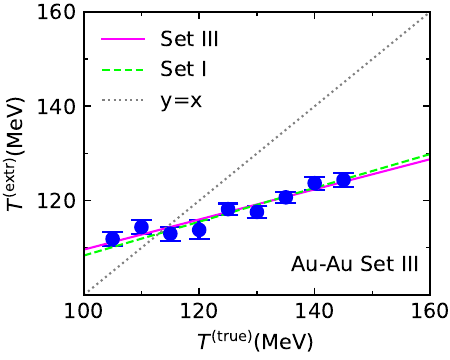}
	\end{minipage}
	\caption{\label{fig:errau3}
		The same as Fig.\ \ref{fig:errau2} for Set III.}
\end{figure}

To further validate the above map, we introduce two more sets of points in the $\tfo^{(\text{true})}$-$(\eta/s)^{(\text{true})}$-plane. We arrive
at these sets by going through Set I and decreasing or increasing, respectively, $\eta/s$ by $0.5/(4\pi)$, keeping the impact parameters and freeze-out temperatures 
in Tab.\ \ref{tab:event}. The resulting sets will be called Set II and III, respectively. Following the same process, we generate fluid dynamic pseudodata for these sets by running MUSIC, and extract $\tfo$ and $\eta/s$ from subsequent blast wave fits of the pseudodata. 
The results are detailed in Appendix \ref{sec:app3}, since we only need to discuss the final conclusion. We find that the extracted values of $\eta/s$ are again consistent with the true values within uncertainties. The general trends seen for the extracted freeze-out temperatures $\tfo$ using Set I are the same
for the two new sets. As a result, the maps we obtain for $\tfo^{(\text{true})} \mapsto \tfo^{(\text{extr})}$ for the new sets are very close to the ones obtained
earlier, as shown in the right panels of Figs.\ \ref{fig:errau2} and \ref{fig:errau3} for the Au+Au case. The Pb+Pb case is analogous.
We conclude that the map for set I already provides a reliable description of the extracted temperatures over a wide range of parameters.

We are now in a position to remove the bias on the extracted temperature that comes from using the approximations of the blast wave. 
By inverting the maps and applying them to the extracted temperatures, we can find "corrected" values, $(\tfo^{(\text{extr})},(\eta/s)^{(\text{extr})}) \mapsto (\tfo^{(\text{corr})},(\eta/s)^{(\text{corr})})$.
The inverted maps are
\begin{align}
\label{eq:corr}
\tfo^{(\text{corr})}= &2.78\tfo^{(\text{extr})}-201.4\, \text{MeV}\qquad \mathrm{(Au+Au)}
\\
\tfo^{(\text{corr})}= &2.13\tfo^{(\text{extr})}-132.8\,  \text{MeV} \qquad \mathrm{(Pb+Pb)}
\end{align}
By virtue of $T_\mathrm{fo}^{  (\text{extr,lin})} \approx T_\mathrm{fo}^{  (\text{extr})}$ we have $T_\mathrm{fo}^{  (\text{corr})} \approx T_\mathrm{fo}^{  (\text{true})}$. Thus we have an algorithm for predicting the correct temperature.
Although our work has focused on freeze-out temperature and specific shear viscosity, the analysis in this work could be repeated for other quantities extracted from blast wave fits in a straight forward way. 

As an application of the above procedure, we point the readers to our extraction of  specific shear viscosity from data in \cite{Yang:2022yxa}. For example, considering the extracted value from ALICE Pb+Pb at 2.76 TeV in the 50-60\% centrality bin we extract the raw values $(T,\eta/s)^{\text{extr}}=(130.1,1.66/(4\pi))$ and subsequently
obtain the corrected values $(T,\eta/s)^{\text{corr}}=(144.3,1.66/(4\pi))$. 
The same process can be applied to other centrality bins of ALICE Pb+Pb as well as RHIC Au+Au at 200 GeV. As a result of the correction $\eta/s$ drops more slowly with increasing temperature, and a value close to $4\pi\eta/s = 1$ is only reached around $T\approx 150$ MeV. 
For further discussions of the implications of this particular result we refer the reader to reference \cite{Yang:2022yxa}. 

\section{Summary}

In this paper we have discussed differences and complementarity between viscous blast wave descriptions and fluid dynamic simulations in the context of high energy nuclear collisions. 
We have carried out a systematic study of parameters at kinetic freeze-out set in fluid dynamic simulations which have subsequently been extracted by blast wave fits to 
hadron spectra and elliptic flow. We find that viscous blast wave fits correctly reproduce broad trends of the true values set in the simulations, and that they qualitatively 
agree with the numerical values set in fluid dynamics. To be more precise, extractions of the specific shear viscosity $\eta/s$ tend to be accurate within expected 
uncertainties. This is backed up a direct comparison of relevant average shear viscosity components between fluid dynamics and blast waves. On the other hand, 
we find the extracted temperatures and average radial flow velocities deviate from their true values. While still somewhat accurate in central collisions, freeze-out
temperatures are underestimated, and average radial flow velocities overestimated in peripheral collisions. However, note that the overall fit quality of hadron spectra 
and elliptic flow remains excellent within the chosen ranges.

The quality of blast wave fits can be improved by understanding and quantifying these deviations. One can establish a map from the true temperature values to the ones extracted from the blast wave fits. The inverse of this map can be applied to arrive at corrected fit values for the temperatures. One can argue that these should be
close to the true values and thus correct the biases in blast wave fits. 
Our study focuses on fitted shear viscosities and temperatures, but it can be readily extended to other physical parameters following the blueprint laid out here. 
The benefit of removing biases in blastwave fits has been demonstrated for the case of the extraction of the specific shear viscosity as a function of 
temperature from data in Ref.\ \cite{Yang:2022yxa}. 
In the latter work, an added complexity is the application of the procedure to experimental data, which leaves the additional question of the quality of hydrodynamic 
modeling of nuclear collisions, as discussed in the original work.


The main point of this paper is the demonstration that average behavior in fluid dynamics is generally well described by viscous blastwaves, and that remaining deviations
can be systematically removed. An obviouly  interesting question emerges regarding the universality and systematic behavior of the specific corrections discussed. 
The maps for freeze-out temperature and shear viscosity discussed here might find direct application if the same blastwave and fit ranges are used for diffent fluid
dynamic simulations. However, caution would dictate that interested readers should always check and if needed establish a custom map for their own setup.
E.g., it is rather obvious that significant changes to the fit ranges, or additional observables, will change the map between true and extracted values.

The viscous blastwave itself could also be improved. The most straight forward changes could be the addition of bulk stress corrections and the inclusion of important hadronic resonances and their decays. The inclusion of resonances has been discussed in \cite{Mazeliauskas:2019ifr}. Indeed, the fits of ALICE data performed in that work lead
to somewhat higher freeze-out temperatures than the raw values found in \cite{Yang:2022yxa}. Qualitatively, this trend is compatiblen with the way the corrections 
found here move raw values. However, a closer inspection of the dependence on fit ranges, which are different in both cases, would be necessary for an qualitative 
comparison.

Finally, let us recall that simplicity is key to blast wave fits. 
While fluid dynamics has matured rapidly and is numerically feasible, blast waves are cheaper in labor and computational requirements. They remain a quick and useful tool for certain data analyses. The formalism laid out here can add significantly to their reliability. 

\begin{acknowledgments}
	ZY would like to thank Mayank Singh and Michael Kordell II for providing MUSIC support. RJF would like to thank Charles Gale and Sangyong Jeon for valuable comments, and Aleksis Mazeliauskas for a useful discussion. This work was supported by the US National Science Foundation under award nos.\ 1516590, 1550221, 1812431 and 2111568 (Z.Y. and R.J.F.), and the National Natural Science Foundation of China under Grants No.\ 12205182, 12235010 and 11625521 (Z.Y.).
\end{acknowledgments}

\begin{appendix}

\section{MUSIC Settings}
\label{sec:app1}

The settings for the fluid dynamic code MUSIC are documented in Tab.\ \ref{tab:setlist} for RHIC energies. Values for LHC are given in parentheses unless they are the same
 as for RHIC.
\begin{table}[htb]
	\centering
	\begin{tabular}{ll}
		\hline
		Parameter & Set\\ \hline\hline
		Target+Projectile    & Au+Au\qquad (Pb+Pb) \\  
		Maximum energy density & 54.0 \qquad (96.0)\footnote{$e_{\textnormal{max}}$= 75.0 was used for $\tfo=110$ MeV}    \\  	
		SigmaNN&42.1   \qquad  (70.0)      \\   
		Initial\_profile & Optical Glauber model\\ 
		EOS\_to\_use &lattice EOS s95p-v1.2 \\ 
	   \hline  	          
		boost\_invariant & 1           \\
		Viscosity\_Flag & 1        \\  
		Include\_Shear\_Visc &1    \\  
		T\_dependent\_Shear\_to\_S\_ratio & 0 \\     
		Include\_Bulk\_Visc &1     \\ 
		Include\_second\_order\_terms &0     \\ 
		Do\_FreezeOut &1        \\ 
		use\_eps\_for\_freeze\_out & use temperature \\ 
		\hline	
		pt\_steps  &36                  \\     
		min\_pt & 0.01             \\          
		max\_pt  &3.0               \\          
		phi\_steps  &40              \\        
		Include\_deltaf\_in Cooper-Frye formula & 1                    \\
		Include\_deltaf\_bulk & 1 \\ 
		\hline
	\end{tabular}
	\caption{\label{tab:setlist} Parameter set for MUSIC runs generating the pseudodata. 1 and 0 are flags corresponding to YES and NO.}
\end{table}

\section{Uncertainty Estimates}
\label{sec:app2}

There are many sources of uncertainties in the extraction of parameters using the viscous blastwave.  
The most important sources have been discussed at the end of Sec.\ \ref{sec:bw}. Some of them can be studied systematically
to assign a measure of the uncertainty to the extracted values of temperature and specific shear viscosity.
In this appendix we investigate three of these uncertainties, which are then used to compute the error bars in Figs.\ \ref{fig:errau} to \ref{fig:errau3}:
(i) uncertainties due to using a fixed value of the radial shape parameter $n$, (ii) uncertainties due to the error bars assigned to MUSIC 
pseudodata, and (iii) uncertainties of the fit itself as encoded in the posterior distributions. 

We have studied the uncertainties from source (i) by systematically varying $n$ by $\pm 0.05$. As an example, we give the results for our Au+Au example parameters
with a true freeze-out temperature of 130 MeV in Tab.\ \ref{tab:errn}. Similarly, we estimate the uncertainty from the error given to the pseudodata by varying that error
by $\pm 1$ percentage point in the analyses. Tab.\ \ref{tab:uncert} gives the results for the same 130 MeV Au+Au example parameter set if this procedure is applied
to the hadron spectrum pseudodata. We find that for theses examples $\tfo$ varies only within a few MeV, while $\eta/s$ shows substantial changes when varying $n$. 
On the other hand, the extracted temperature and specific shear viscosity are largely insensitive to variations of the assigned pseudodata error.

\begin{table}[htb]
	\centering
	\begin{tabular}{|l|l|l|l|}
		\hline
		  & $n$ & $\tfo$(MeV)& $4\pi \eta/s$  \\
		\hline
		small & 0.83 & 116.4 & 3.26  \\
		\hline
		regular & 0.88 & 116.4 & 2.47\\
		\hline
		large & 0.93 & 117.8 & 1.81\\
		\hline
	\end{tabular}
	\caption{\label{tab:errn}The extracted values of $\tfo$ and $\eta/s$ for different values set for $n$ in the viscous blastwave. This example uses MUSIC Au+Au pseudodata for $\tfo$ =130 MeV.}
\end{table}

\begin{table}[htb]
	\centering
	\begin{tabular}{|l|l|l|l|}
		\hline
		 & spectra uncertainty & $\tfo$(MeV)& $4\pi \eta/s$\\
		\hline
		small &4\%  & 116.2 & 2.43 \\
		\hline
		regular &5\%  & 116.4 & 2.47 \\
		\hline
		large&6\%  & 117.8 & 2.62 \\
		\hline
	\end{tabular}
	\caption{\label{tab:uncert}The extracted values of $\tfo$ and $\eta/s$ for different uncertainties assigned to MUSIC pseudodata hadron 
  spectra for Au+Au with $\tfo$=130 MeV. The uncertainty for $v_2$ is fixed at 2\% with a pedestal 0.002. The uncertainty for spectra is varied 
  as shown in the table.}
\end{table}

Uncertainties of type (iii) can be taken directly from the MADAI output. We treat the sources of uncertainties as independent and add them quadratically to arrive at 
estimates of the total uncertainty assigned to the extracted values of $\tfo$ and $\eta/s$. For the point selected for this example these errors are summarized in Tab.\ \ref{tab:uncertAll} for the 130 MeV Au+Au case. We repeat the uncertainty estimates for the remaining points of sets I through III which provides the error bars in
Figs.\ \ref{fig:errau} through \ref{fig:errau3}.

\begin{table}[htb]
	\centering
	\begin{tabular}{|l|l|l|l|l|}
		\hline
		Origin of uncertainty  & $n$ &error assigned& stat.\ analysis & total $\sigma$ \\
		\hline
		$\sigma_T$ (MeV) & 0.66 & 0.71 & 1.47 & 1.76 \\
		\hline
		$\sigma_{\eta}(4\pi)$  & 0.59 &0.08 & 0.22 & 0.64  \\
		\hline
	\end{tabular}
	\caption{\label{tab:uncertAll} A summary of uncertainties for temperature and specific shear viscosity extracted from MUSIC pseudodata for Au+Au with $\tfo$=130 MeV.}
\end{table}

\section{Fits for Sets II and III and Uncertainty of the Mapping}
\label{sec:app3}

To validate the linear maps between the extracted and true temperatures and shear viscosities, we test two additional sets of points in the $\tfo^{(\text{true})}$-$(\eta/s)^{(\text{true})}$-plane for fluid dynamics simulations of Au+Au collisions by decreasing or increasing $\eta/s$ by $0.5/(4\pi)$ with the same impact parameters and 
freeze-out temperatures given in Tab.\ \ref{tab:event}. These are called Sets II and III, respectively.
Following the same process as before, we generate MUSIC pseudodata and extract $\tfo$ and $\eta/s$ from blast wave fits of the hadron spectra and elliptic flow. The 
true and extracted values are listed in Tabs.\  \ref{tab:event2} and \ref{tab:event3}. 

\begin{table}[tbh]
	\caption{\label{tab:event2} Set II of parameters for running MUSIC simulations of Au+Au collisions, using the same impact parameters and freeze-out temperatures as Au+Au collisions in Tab.\ \ref{tab:event} but smaller $\eta/s$. The raw ${\tfo}$ and $\eta/s$ values extracted from the blastwave are also shown. }
	\centering
	\begin{tabular}{|l|l|l|l|l|l|l|l|l|l|l|}
		\hline
		\multirow{2}*{Hydro (true)} & $\tfo$ (MeV) & 105 & 110 & 115 & 120 & 125 & 130 &  135 & 140 & 145 \\
		\cline{2-11}      
		&$4\pi\eta/s$  & 5.28& 4.52& 3.77& 3.02& 2.51&  2.01& 1.51& 1.01 &0.05 \\
		\hline
		\multirow{2}*{Blastwave (extr)} & $\tfo$ (MeV)  & 110.7 & 113.8 & 111.9 & 113.4 & 118.5 & 117.5 & 121.4 & 124.6& 128.4 \\
		\cline{2-11}      
		&$4\pi\eta/s$  &5.23 &4.79 &3.73 &3.14 &2.84 & 2.36 & 1.55 & 1.10 &0.60 \\
		\hline
	\end{tabular}
\end{table}

\begin{table}[tbh]
	\caption{\label{tab:event3} Set III of parameters for running MUSIC Au+Au simulations, using the same impact parameters and freeze-out temperatures as Au+Au collisions in Tab.\ \ref{tab:event} but larger $\eta/s$. The ${\tfo}$ and $\eta/s$ values extracted from the blastwave are also shown. }
	\centering
	\begin{tabular}{|l|l|l|l|l|l|l|l|l|l|l|}
		\hline
		\multirow{2}*{Hydro (true)} &$\tfo$ (MeV) & 105 & 110 & 115 & 120 & 125 & 130 &  135 & 140 & 145 \\
		\cline{2-11}      
		&$4\pi\eta/s$ & 6.79 & 6.03&  5.28& 4.52& 3.77& 3.02& 2.51&  2.01& 1.51 \\
		\hline
		\multirow{2}*{BlastWave (extr)} & $\tfo$ (MeV)  & 111.9 & 114.4 & 113.0 & 113.8 & 118.2 & 117.6 &  120.7 & 123.7 & 124.4 \\
		\cline{2-11}      
		&$4\pi\eta/s$  &6.50 &6.11 &5.21 &4.06 &3.62 & 2.78 & 2.55 & 1.80 &1.44 \\
		\hline
	\end{tabular}
\end{table}

Once again, the values for $\eta/s$ extracted from the blastwave fits are consistent with the true values within uncertainty estimates.
The extracted freeze-out temperatures follow trends very much similar to before. We can again obtain an approximation $\tfo^{(\text{extr,lin})}$ to the extracted 
temperatures by linear regression for each set. For Set II, we obtain $\tfo^{(\text{extr,lin})}=0.41\tfo^{(\text{true})}+66.3\, \text{MeV}$. For Set III, we have 
$\tfo^{(\text{extr,lin})}=0.32\tfo^{(\text{true})}+77.7\, \text{MeV}$. The correlations between true and extracted values of the temperature, together with the
new linear maps are shown in Figs.\ \ref{fig:errau2} and \ref{fig:errau3}. We also show the map for Set I in each case for comparison. Clearly, the original map is
consistent with Sets II and III. We can thus declare it sufficiently accurate in a region of the $\tfo$-$(\eta/s)$-plane around the original Set I with a width of $\sim 1$
in $4\pi\eta/s$.


\end{appendix}

\bibliography{ref} 
\include{ref}

\end{document}